\documentstyle[11pt,aaspp4]{article}

\def\etal {\emph{et~al.\hspace{.1in}}}

\def\go {\mathrel{\raise.3ex\hbox{$>$}\mkern-14mu\lower0.6ex\hbox{$\sim$}}}
\def\lo {\mathrel{\raise.3ex\hbox{$<$}\mkern-14mu\lower0.6ex\hbox{$\sim$}}}
\def\fd {\dot f}
\def\fdd {\ddot f}
\def\fddd {\stackrel {\ldots}{f}}
\def\fdddd {\stackrel {\ldots.}{f}}
\def\fddddd {\stackrel {\ldots..}{f}}

\begin{document}

\title{Distant Companions and Planets around Millisecond Pulsars}

\author{Kriten J.~Joshi\altaffilmark{1} and Frederic A.~Rasio\altaffilmark{2,3}}

\affil{Department of Physics, Massachusetts Institute of Technology}

\altaffiltext{1}{6-218M MIT, 77 Massachusetts Ave, Cambridge, MA 02139; email: kjoshi@mit.edu.}
\altaffiltext{2}{6-201 MIT, 77 Massachusetts Ave, Cambridge, MA 02139; email: rasio@mit.edu.}
\altaffiltext{3}{Alfred P.\ Sloan Foundation Fellow.}

\begin{abstract}
We present a general method for determining the masses and
orbital parameters of binary millisecond pulsars with long
orbital periods ($P_{\rm orb} \gg 1\,$yr), using timing data in
the form of pulse frequency derivatives.
Our method can be used even when the available
timing data cover only a small fraction of an orbit, but it requires high-precision
measurements of up to five successive derivatives of the pulse frequency.
With five derivatives 
a complete determination of the mass and orbital parameters is in principle possible
(up to the usual inclination factor $\sin i$). With less than five derivatives
only partial information can be obtained, but significant constraints can sometimes
be placed on, e.g., the mass of the companion.
We apply our method to analyze the properties of the second
companion in the PSR~B1620-26 triple system.
We use the latest timing data for this system, including
a recent detection of the fourth time derivative of the pulse
frequency, to constrain the mass and orbital parameters of the second companion.
We find that all possible solutions have a mass $m_2$ in the range
$2.4\times10^{-4}\,\rm M_\odot \leq m_2 \sin i_2 \leq 1.2\times10^{-2}\,\rm M_\odot$,
i.e., almost certainly excluding a second companion of stellar mass and suggesting
instead that the system contains a planet or a brown dwarf.
To further constrain this system we have used preliminary measurements of the 
secular perturbations of the inner binary. Using
Monte-Carlo realizations of the triple configuration in three dimensions
we find the most probable
value of $m_2$ to be $0.01\pm0.005\,\rm M_\odot$, corresponding to
a distance of $38\pm6\,\rm AU$ from the center of mass of the inner binary 
(the errors indicate 80\% confidence intervals).
We also apply our method to analyze the planetary system around
PSR~B1257$+$12, where a distant, giant planet may be present in
addition to the three well-established Earth-mass planets.
We find that the simplest interpretation of the frequency derivatives 
implies the presence of a fourth planet
with a mass of  $\sim100\,\rm M_\oplus$
in a circular orbit of radius $\sim40\,\rm AU$.

\end{abstract}


\section{Introduction}

The traditional method for obtaining masses and orbital parameters
of binary pulsars consists of fitting Keplerian or Post-Newtonian
models to timing data covering at least one complete orbit.
For wide-orbit binary pulsars, with orbital periods longer than a
few years to decades, fitting a complete orbit may not be possible.
For such systems, we present a method for obtaining the masses and
orbital parameters using measured values of the successive time
derivatives of the pulse frequency ($\fd$, $\fdd$, $\fddd$, etc.). 
Given the high precision of millisecond pulsar timing
it is sometimes possible to measure these frequency derivatives
up to high order with only a few years of observations.
We show below that with {\em five derivatives\/} a complete solution may be 
obtained, with the orbital parameters and companion mass fully determined
up to the usual unknown inclination factor $\sin i$. 
This solution is constructed under the assumption that all frequency derivatives
are {\em dynamically induced\/} rather than being intrinsic to the pulsar
spin-down. This is a crucial assumption which may not always be justified.
Its validity, and the effects of relaxing it, must be examined for each
particular application. If only three or four derivatives have been
measured, significant constraints can still be placed on the
parameters of the system. With just three dynamically-induced derivatives and 
the assumption of a circular orbit (often justified, e.g., for a planet)
a complete solution can again be obtained.
Our method is not a substitute for the standard fitting procedure to data covering
a complete orbit. Instead, it provides a way of obtaining the orbital 
parameters of wide-orbit binaries which cannot be observed over a complete 
orbit due to their very long orbital periods. The method can only 
be successfully applied to binaries containing fast millisecond pulsars
with low timing noise, in which one can reasonably expect the dynamically-induced
pulse frequency derivatives to dominate over intrinsic changes.

Applications of our method to two systems, PSR~B1620$-$26
and PSR~B1257$+$12, will be presented in this paper. Both of these systems
are thought to contain planetary-mass companions to the millisecond pulsar
(Wolszczan 1994; Arzoumanian \etal 1996). 
Their properties are of great interest
for our understanding of planet formation outside the solar system.
This is particularly true in light of
the many recent detections of extrasolar planets around
nearby stars, which show a great diversity of properties (Mayor \& Queloz 1995; 
Butler \& Marcy 1996; Marcy \& Butler 1996).
PSR~B1620$-$26, in the globular cluster M4, is in a hierarchical triple 
configuration. Previously available timing data allowed a second companion mass 
anywhere in the range $\sim10^{-3}$--$1\,\rm M_\odot$ (Michel 1994;
Sigurdsson 1995), i.e., including the possibility of a Jupiter-type planet.
PSR~B1257$+$12 has three confirmed low-mass planets, and there is recent evidence for 
a fourth more massive one (Wolszczan 1996).

Our paper is organized as follows. In \S 2, we describe
our general method for obtaining the companion mass and the
orbital parameters using measured pulse frequency derivatives.
In \S 3, we apply our method to PSR~B1620$-$26.
We also incorporate into our analysis preliminary measurements
of secular perturbations of the inner binary by
performing Monte-Carlo simulations with the undetermined
parameters in the system to obtain the most probable mass for the
second companion.
In \S 4, we apply our method to check whether the observed
frequency derivatives of PSR~1257$+$12 are consistent with the
possible existence of a fourth object in the planetary system.

\section{Inverting Frequency Derivative Data}

In the standard method for determining the parameters
of binary pulsars, the data (consisting of radio pulse arrival times)
are fitted to the predictions of a Keplerian or
Post-Newtonian model of the binary orbit. To obtain a reliable fit, one usually
needs timing data covering at least a few complete orbital periods.
However, even if the data do not cover a full period, it is still 
possible to determine, at least approximately,
the companion mass and orbital parameters of the system 
if sufficiently accurate timing data are available.
The method we develop here uses time 
derivatives of the pulse frequency (basically the coefficients 
in a Taylor expansion of the pulse frequency around a particular epoch).
Pulse frequency derivatives are a convenient way in
which radio astronomers can present the results of their observations
when a clear periodicity cannot be recognized in the timing data. 

\subsection{General Formulation}

Assuming that the pulsar mass ($m_1\simeq1.4\,M_\odot$)
is known, there are five parameters
that can in principle be determined using our method: the mass of the 
companion $m_2$ (up to the unknown inclination angle $i_2$), the
semi-major axis $a_2$, the eccentricity $e_2$, the longitude of 
pericenter $\omega_2$
(measured from the ascending node) and the longitude $\lambda_2$
of the companion at the reference epoch (measured from 
pericenter). 
The inclination angle $i_2$ is the angle between the normal to the orbital 
plane and the line of sight; this angle cannot be determined directly 
from the timing data. Here and throughout this paper a subscript~1 refers to
the pulsar, a subscript~2 refers to the companion, and all orbital elements
correspond to the motion around the center of mass of the system (e.g., the
distance between the pulsar and its companion is $r_1+r_2$). 

If frequency derivatives up to the {\em fifth order\/} are 
measured, all five parameters ($m_2$, $a_2$, $e_2$, $\omega_2$, $\lambda_2$)
can in principle be determined. For small values of $m_2/m_1$ the usual 
combination $m_2 \sin i_2$ can be obtained 
(see the discussion for PSR~B1620$-$26 in \S3). For larger companion masses,
the dependence of the solutions on $i_2$ is more complicated and  
one needs to adopt a particular value of $\sin i_2$ (in practice $\sin i_2\lo1$ for
a random orientation) in order to solve explicitly for $m_2$.

Our method assumes that the measured frequency derivatives are purely those 
induced by the motion of the pulsar (acceleration, jerk, and higher derivatives)
around the binary center of mass. This requires correcting the measurements
for other possible kinematic effects (see, e.g., Camilo \etal 1994). 
More importantly, one must assume that
any intrinsic contribution from the pulsar spin-down can be neglected, or
determined to some extent from the known properties of other millisecond 
pulsars and subtracted from the measured values.
In particular, the observed first 
frequency derivative $\fd$ is in general determined by
a combination of acceleration and intrinsic spin-down of the pulsar, 
\begin{equation}
\fd_{\rm obs} = \fd_{\rm int} + \fd_{\rm acc}.
\end{equation}
It is not possible to measure the two components separately in general. 
For some systems, however, it is reasonable to 
assume that $|\fd_{\rm int}| \ll |\fd_{\rm acc}|$ if $|\fd_{\rm obs}|$ is large.
For example, the observed $\fd_{\rm obs}$ may be positive, a clear sign that
it is determined predominantly by acceleration 
(as in the M15 pulsars PSR 2127+11~A and~D; Wolszczan et al.\ 1989). 
The expected value of $\fd_{\rm int}$ may also be estimated from the pulse 
frequency $f$ and from the assumption that the timing age $\tau\equiv f/(2\fd)$
of a millisecond pulsar should satisfy $\tau\go 10^9\,\rm yr$. Indeed most
millisecond pulsars with reliably measured timing ages appear to
satisfy this property (e.g., Phinney \& Kulkarni 1994). 
Note that the true ages of some millisecond pulsars
may be considerably smaller (cf.\ Lorimer \etal 1995), but this does not 
affect our argument.
Similarly, the expected value of $\fdd_{\rm int}$ can be estimated from
the predicted level of {\em timing noise\/} for the pulsar, which, 
although very small for millisecond pulsars, may also affect the interpretation of 
$\fdd_{\rm obs}$ (see, e.g., Arzoumanian \etal 1994; Kaspi \etal 1994). 
Intrinsic higher derivatives ($\fddd$, etc.) are normally not 
measurable for millisecond pulsars, and therefore we can always assume 
safely that any measured values are dynamically induced.

After subtraction of known kinematic and intrinsic contributions,
we can write the  time derivatives of the pulse
frequency at a particular reference epoch as 
\begin{eqnarray}
\fd & = & -f {{\bf a}\cdot{\bf n}\over c}, \\
\fdd & = & -f {{\dot{\bf a}}\cdot{\bf n}\over c} + {{\fd}^2\over f}, 
\rm\, etc., \nonumber
\end{eqnarray}
where $c$ is the speed of light,  $\bf a$ is the acceleration of the
pulsar, and
${\bf n}$ is a unit vector in the direction of the line of sight,
and a dot indicates a time derivative.
The second term in the expression for $\fdd$ is smaller than the first
by a factor $\sim v/c$, where $v$ is the orbital velocity of the
neutron star, i.e.,
${\fd}^2/f\ll |{\fdd}|$. For $\fddd$, we have
$|{\fd}{\fdd}/f| \ll |{\fddd}|$, etc., so that all similar terms can be neglected
in taking higher and higher derivatives.
Therefore, we can write the first five frequency derivatives 
simply as
\begin{eqnarray}
\fd & = & -f {{\bf a}\cdot{\bf n}\over c}, \nonumber \\
\fdd & = & -f {{\dot{\bf a}}\cdot{\bf n}\over c}, \\
 &{\rm \vdots}& \nonumber \\
\fddddd & = & -f {{\stackrel{\ldots.}{\bf a}}\cdot{\bf n}\over c}.\nonumber
\end{eqnarray}
Equations (3) form a system of five nonlinear algebraic equations with five 
unknowns ($m_2$, $a_2$, $e_2$, $\lambda_2$, and $\omega_2$) which must
be solved numerically. It is straightforward but tedious to write down
explicitly the right-hand sides of these equations in terms of the five unknowns,
and we will omit their explicit forms in the general case.
Some of the steps involved, however, are described below in \S2.2.

\subsection{Solution with Four Derivatives}

When four derivatives ($\fd$ through $\fdddd$) are known, one can obtain 
a one-parameter family of solutions. In practice, the problem 
can be reduced to solving a system of {\em three\/} nonlinear equations as follows.
If $m_1$ is the mass of the pulsar, and $m_2$ 
is the mass of the companion, the acceleration of the pulsar in its motion around
the center of mass of the binary is of magnitude $a = k/r_1^2$, where $r_1$ is the
distance from the pulsar to the center of mass and
\begin{equation}
k = G {m_2^3\over (m_1+m_2)^2}.
\end{equation}
For an elliptic Keplerian orbit of semimajor axis $a_1$ and eccentricity $e_1=e_2$, 
the distance $r_1$ is given by
\begin{equation}
{1\over r_1} = {1\over h} (1 + e_2\cos\lambda_1) \equiv {A \over h},
\end{equation}
where $h = a_1 (1-e_2^2)$. 
Equations~(3) for the pulse frequency derivatives can then be written 
\begin{eqnarray}
\fd = -f{a\over c} \sin(\lambda_1+\omega_1) \sin i_2
& = & -f K A^2 \sin(\lambda_1+\omega_1),\\
\fdd & = & -f K B \dot\lambda_1, \\
\fddd & = & -f K C \dot\lambda_1^2, \\
{\rm and}\hspace{0.5in}
\fdddd & = & -f K D \dot\lambda_1^3,
\end{eqnarray}
where  we have defined
\begin{eqnarray}
A & = & 1 +e_2\cos\lambda_1, \nonumber \\
B & = & 2AA'\sin(\lambda_1+\omega_1)+A^2\cos(\lambda_1+\omega_1),\nonumber \\
C & = & B' +{2BA' \over A},  \\ 
D & = & C' +{4CA'\over A}, \nonumber \\ {\rm and}\hspace{0.5in}
K & = & {k \sin i_2 \over h^2 c}, \nonumber
\end{eqnarray}
and a prime indicates a derivative with respect to $\lambda_1$.

Using equation (6), we can rewrite equations (7)--(9) as
\begin{eqnarray}
\fdd & = &{B{\dot\lambda_1}\fd\over A^2\sin(\lambda_1+\omega_1)}, \\
\fddd & = &{C{\dot\lambda_1}^2\fd\over A^2\sin(\lambda_1+\omega_1)}, \\
{\rm and}\hspace{0.5in}
\fdddd & = & {D{\dot\lambda_1}^3\fd\over A^2\sin(\lambda_1+\omega_1)}.
\end{eqnarray}
This is a nonlinear system of three equations with four unknowns- 
$e_2$, $\lambda_1=\lambda_2$, $\omega_1=\omega_2 + 180^{\circ}$, 
and $\dot\lambda_1$. Assuming a value 
for one of them we can solve for the remaining parameters. 

We have chosen to use the eccentricity $e_2$ as our free parameter.
For an assumed value of $e_2$, we solve equations (11)--(13)
for $\lambda_1$, $\dot\lambda_1$, and $\omega_1$ using
the Newton-Raphson method (see e.g., Press \etal 1992).
This method requires
an initial guess for the unknown parameters, which is then successively
improved until convergence to an actual solution is obtained. One must
be careful to experiment with many different initial guesses, since
nonlinear systems like this one  often have multiple branches of solutions.
In addition, symmetries 
must also be taken into account. Here physically equivalent 
solutions are obtained if the direction of motion is reversed, and the 
signs of $\omega_1$ and $\lambda_1$ are also reversed. This gives a different 
but equivalent orientation of the system. 
Once $\lambda_1$, $\dot\lambda_1$, and $\omega_1$ are known, 
conservation of angular momentum together with 
equation~(6) gives
\begin{eqnarray}
\frac{k}{h^3} & = & \frac{\dot\lambda_1^2}{A^4}, \\ {\rm and}\hspace{.5in}
\frac{k}{h^2} & = & -\frac{\fd c}{f A^2 \sin(\lambda_1+\omega_1) \sin i_2}.
\end{eqnarray}
Using equations (14) and~(15), we can calculate $k$ and $h$,
\begin{eqnarray}
h & = & -\frac{\fd c A^2}{f \sin i_2 \sin \lambda_1 \dot\lambda_1^2}, \\
{\rm and}\hspace{0.5in}
k & = & -\left(\frac{\fd c}{f \sin i_2 \sin \lambda_1}\right)^3 
\left(\frac{A^2}{\dot\lambda_1^4}\right).
\end{eqnarray}
It is now straightforward to obtain $m_2$ (from $k$, assuming that the
pulsar mass $m_1$ is known) and the
semi-major axis $a_2=(m_1/m_2)h/(1-e_2^2)$.
When $m_2 \ll m_1$, we can directly obtain $m_2 \sin i_2$ as
follows. For small $m_2$, $k \approx G m_2^3/m_1^2$.
Therefore $m_2 \approx (k m_1^2 / G)^{1/3}$ and equation~(17) gives
\begin{eqnarray}
m_2 \sin i_2 \approx \frac{\fd c}{f \sin \lambda_1}\left(\frac{m_1^2 A^2}
{G \dot\lambda_1^4}\right)^{1/3}.
\end{eqnarray}
For larger values of $m_2/m_1$, one needs to assume a value for $\sin i_2$ and 
solve equation~(4) numerically for $m_2$.

\subsection{Solution with Three Derivatives}

When only three derivatives are known, one can assume values for two 
parameters and solve for the remaining three parameters in the same way
as above. 
Alternatively, in the special case of a circular orbit ($e_2= 0$), 
the system can be solved 
completely (to within $\sin i_2$) using only three derivatives. In that 
case, we have
\begin{eqnarray}
\fd & = & \frac{-f k \sin i_2}{a_1^2\,c} \sin(\lambda_1), \\
\fdd & = & \frac{-f k \sin i_2}{a_1^2\,c} \cos(\lambda_1) \dot \lambda_1, \\
{\rm and} \hspace{0.5in}
\fddd & = & \frac{f k \sin i_2}{a_1^2\,c} \sin(\lambda_1) \dot\lambda_1^2.
\end{eqnarray}
Note that here $\lambda_1$ is the longitude of the pulsar {\em measured from
the ascending node\/} ({\em not\/} from pericenter as before).

Using equation~(19) to eliminate $-f k \sin i_2/a_1^2\,c$, we get
\begin{eqnarray}
\fdd & = & \frac{\fd}{\sin(\lambda_1)} \cos(\lambda_1) \dot \lambda_1, \\
{\rm and}\hspace{.5in} 
\fddd & = & - \fd \dot\lambda_1^2.
\end{eqnarray}
This gives
\begin{eqnarray}
\dot\lambda_1^2  & = &  - (\fddd / \fd), \\ {\rm and}\hspace{.5in}
\lambda_1 & = & \arctan\left[\frac{\fd}{\fdd}
		\left(\frac{-\fddd}{\fd}\right)^{1/2}\right].
\end{eqnarray}
Using equation~(19) and conservation of angular momentum, we have
\begin{eqnarray}
\frac{k}{a_1^2} & = & \frac{-\fd c}{f \sin i_2 \sin(\lambda_1)}, \\
{\rm and}\hspace{.5in}
\frac{k}{a_1^3} & = & \dot\lambda_1^2.
\end{eqnarray}
Dividing equation (26) by equation~(27), and then using equations (24), (25) and~(27),
we get
\begin{eqnarray}
a_1 & = & \frac{\fd^2 c}{f \fddd \sin i_2 \sin(\lambda_1)}, \\
{\rm and}\hspace{.5in}
k   & = & -\left(\frac{\fd c}{f \sin i_2 \sin(\lambda_1)}\right)^3
	\left(\frac{\fd}{\fddd}\right)^2,
\end{eqnarray}
where $\lambda_1$ is given by equation (25).
For $m_2 \ll m_1$, $k \approx G m_2^3/m_1^2$. Then, for given
values of the frequency derivatives and the pulsar mass $m_1$, 
$m_2$ can be calculated explicitly using equation~(29). We find
\begin{eqnarray}
m_2 \sin i_2 & \approx & -\left(\frac{m_1^2}{G}\right)^{1/3}
\left(\frac{\fd c}{f \sin(\lambda_1)}\right) \left(\frac{\fd}{\fddd}\right)^{2/3},
\end{eqnarray}
where $\lambda_1$ is given by equation~(25). Finally we can calculate
$a_2 = (m_1/m_2) a_1$, where $a_1$ is given by equation~(28).

\subsection{Solution with Two Derivatives}

When only two derivatives are available, one can obtain a one-parameter
family of solutions, assuming again that $e_2 \approx 0$.
We can use $\fddd$ (or 
equivalently $\dot\lambda_1$, using eq.~[24]) as the free parameter 
and then use the equations of 
\S 2.3 to construct a one-parameter family of solutions explicitly.

\section{Application to the PSR~B1620$-$26 Triple System}

The millisecond pulsar PSR~B1620$-$26, in the globular cluster M4,
has a low-mass binary companion (probably a white dwarf of mass
$m_c\approx0.3\,\rm M_\odot$ for a pulsar mass $m_p=1.35\,\rm M_\odot$)
in a $191\,$day low-eccentricity orbit (Lyne \etal 1988; McKenna \& Lyne
1988). The unusually large
frequency second and third derivatives indicate the presence of a
{\em second companion\/} around the inner binary, forming a hierarchical
triple configuration (Backer 1992; Backer, Foster, \& Sallmen 1993; 
Thorsett, Arzoumanian, \& Taylor 1993).
Such hierarchical triple systems are expected to be produced
quite easily in dense globular clusters through dynamical interactions
between binaries. In a typical interaction, one 
star would be ejected, leaving the other three in a stable triple
system (Rasio, McMillan, \& Hut 1995; Sigurdsson 1995).
Previous calculations (done using frequency derivatives up to the
third order) showed that the mass of the second companion
could be anywhere from $\sim 10^{-3}\,\rm M_\odot$ to $\sim 1\,\rm M_\odot$ 
(Michel 1994; Sigurdsson 1995). Recently, a measurement has been made of the
fourth derivative of the pulse frequency, along with preliminary
measurements of secular changes of the inner binary parameters due to
the perturbation of the second companion (Arzoumanian \etal 1996). 
These include a precession of the inner binary orbital plane 
(measured as a change in the projected semi-major
axis of the binary), and possible changes in the eccentricity and 
longitude of periastron.

\subsection{Modeling the Frequency Derivatives}

In this section, we apply the method described in \S 2.2 to analyze the
properties of the second companion in the PSR~B1620$-$26 triple
system. Since the orbital period of the second companion is much
longer than that of the inner binary (for all solutions obtained below),
we treat the inner binary as a single object. Keeping the same notation as
before, we let $m_1 = m_p + m_c$ be the mass of the inner binary pulsar, with
$m_p$ the mass of the neutron star and $m_c$ the mass of the (inner) companion,
and we denote by $m_2$ the mass of the second companion (to be determined).
As in \S2 the orbital parameters ($\lambda_2$, $\omega_2$, $e_2$, $a_2$ and $i_2$)
refer to the orbit of the (second) companion with respect to the center of 
mass of the system (here the entire triple). However, a subscript 1 for the 
orbital elements refers to the orbit of the inner binary.
The results presented in this section are all for $m_1=1.7\,M_\odot$
(assuming $m_p=1.4\,M_\odot$, $m_c=0.3\,M_\odot$; cf.\ Thorsett \etal 1993). 
However, we have have checked that  
they are not very sensitive to small changes in the value 
of $m_1$. In particular the companion mass $m_2$ varies only as $\sim m_1^{2/3}$
(cf. eq.~[4]).

We use the latest available values of the pulse frequency derivatives
(Arzoumanian \& Thorsett 1996) for the epoch MJD 48725.0:
\begin{eqnarray}
f & = & 90.2873320054(1) \, {\rm s}^{-1} \nonumber \\
\fd & = & -5.4702(7)\times10^{-15}\,  {\rm s}^{-2} \nonumber \\
\fdd & = & 1.929(8)\times10^{-23} \, {\rm s}^{-3} \\
\fddd & = & 8(1)\times10^{-33} \, {\rm s}^{-4} \nonumber \\
\fdddd & = & -2.1(6)\times10^{-40} \, {\rm s}^{-5} \nonumber
\end{eqnarray}
These values take into account a (very precise) Keplerian model of the
inner orbit. The corrections to $\fd$ due to proper motion are negligible
for this pulsar. The frequency derivatives 
should therefore reflect the residual motion of the pulsar caused by the
presence (unmodeled) of a second companion. However,
as discussed in \S2.1, the observed first derivative $\fd_{\rm obs}$ can
be a combination of the intrinsic spin-down of the pulsar
and the acceleration due to the second companion.
Since $\fd_{\rm obs}$ is negative, and $\fd_{\rm int}$ is always negative,
$\fd_{\rm acc}$ can in principle be either positive or negative. If it is negative,
then its magnitude must be $\leq \mid\fd_{\rm obs}\mid$. If it is
positive, its magnitude can in principle be larger. 
However, it cannot be much
larger than $\mid\fd_{\rm obs}\mid$ since this would imply a very large
intrinsic spin-down rate, and a short characteristic age
$\tau = -f/(2\dot f)$, which is not expected for a millisecond pulsar 
(cf.\ Thorsett \etal 1993 and \S 2.1).
In practice, we find that varying $\fd_{\rm acc}/\fd_{\rm obs}$ in the entire
range $-1.0$ to $+1.0$ does not affect our solutions significantly 
(see below).
The expected value of $\fdd$ due to timing noise can be estimated 
using, e.g., Figure~1 of Arzoumanian \etal (1994), which gives the timing
noise parameter $\Delta_8 \equiv \log(|\fdd|10^{24}/6f)$ as a function of period
derivative. This gives an {\em upper limit\/} on the contribution to 
$\fdd$ due to intrinsic timing noise of
$3\times10^{-24}\,{\rm s^{-3}}$, which is an order of 
magnitude smaller than $|\fdd_{\rm obs}|$.
The same conclusion is reached if we consider for comparison PSR~B1855+09, 
which has a comparable spin rate
($f = 186\,{\rm s}^{-1}$) but a frequency second derivative
$|\fdd_{\rm obs}| \leq 2\times10^{-27} {\rm s}^{-3}\,(3 \sigma)$ i.e., at 
least four orders of magnitude smaller than $\fdd_{\rm obs}$ in 
PSR~B1620-26 (Kaspi \etal 1994). 

Figure~1 illustrates our ``standard solution'', obtained 
under the assumption that $\fd_{\rm acc} = \fd_{\rm obs}$ and using the 
current best-fit value for the fourth derivative,
$\fdddd_m = -2.1\times10^{-40} \, {\rm s}^{-5}$. Following the method of
\S 2.2, we use $e_2$ as the free parameter. We see that there are no
solutions for $e_2  \lo 0.1$. Hence a nearly circular orbit is ruled out.
For $0.1 \lo e_2 \lo 0.3$ there are two solutions for each value of the
eccentricity, and hence two possible values of $m_2$. 

In the first branch of solutions,
$m_2$ approaches zero as the eccentricity approaches the value 
$e_2 \approx 0.33$. However for very small $m_2$ the second
companion gets close enough to the inner binary to make 
the triple configuration dynamically unstable. The stability of 
the triple system can be checked using an approximate criterion of the
form $Y \geq Y_{\rm min}$ (for stability), where 
$Y = [(1-e_2)(a_1+a_2)]/[(a_p+a_c)(1+e_1)]$ is the
ratio of the outer pericenter separation (from the center of mass
of the inner binary to the second companion) to the apastron separation 
of the inner binary.
Here we have used the results of
Eggleton \& Kiseleva (1995), who give 
\begin{eqnarray}
Y_{\rm min} &  \approx & 1 + {3.7\over q_{\rm out}^{1/3}} + {2.2\over 
{1 + q_{\rm out}^{1/3}}} + {1.4\over q_{\rm in}^{1/3}} 
{{q_{\rm out}^{1/3} -1} 
\over {q_{\rm out}^{1/3} + 1}}.
\end{eqnarray}
Here $q_{\rm in} = m_p/m_c$ and $q_{\rm out} = m_1/m_2$ 
are the inner and outer mass ratios.
Using the values $m_p \approx 1.4\,\rm M_\odot$, $m_c \approx 
0.3\,\rm M_\odot$,
and $m_2 \sim 10^{-5}\,\rm M_\odot$, we get $Y_{\rm min} \approx 2$.
Thus we require that $Y \geq 2$ for a dynamically stable solution.
This gives us a lower limit on the second companion mass 
$m_2 \go 3\times10^{-5}\,{\rm M_\odot} \approx 10\,{\rm M_\oplus}$.
In addition, we can rule out solutions with orbital 
periods $P_2 \lo 14\,\rm yr$ since this would have been detected already 
in the timing residuals (S.~E.\ Thorsett, personal communication). 
This gives us a somewhat stricter lower limit of 
$m_2 \go 2.4\times10^{-4}\,{\rm M_\odot} \approx 80 {\rm M_\oplus}$.

For the second branch of solutions, $m_2$ increases monotonically from
$\sim10^{-3}\,\rm M_\odot$ to $1.2\times10^{-2}\,\rm M_\odot$
(i.e., Jupiter to brown-dwarf masses) as $e_2$ increases from $\sim0.1$ 
to $1$. As $e_2$ approaches 1.0, the mass $m_2$ remains bound. 
So even though the solutions with $e_2 \rightarrow 1$ 
are a priori unlikely, they provide a strict upper limit 
$m_2 \leq 1.2\times10^{-2}\,\rm M_\odot$.
Since all our solutions  have $m_2 \ll m_1$, 
our method gives in fact the product $m_2 \sin i_2$ (cf. \S 2.2), which 
we show in Figure~1.

The effect of varying $\fd_{\rm acc}/\fd_{\rm obs}$ in the 
range 0--1.0 is shown in Figure~2. 
We see that smaller values of
$\fd_{\rm acc}/\fd_{\rm obs}$ give slightly smaller mass solutions,
but the overall mass range is not significantly affected.
In particular, we see that stellar-mass 
solutions for $m_2$ are
not allowed even if the observed value of $\fd$ is assumed to be mostly
intrinsic. 
Negative values of $\fd_{\rm acc}/\fd_{\rm obs}$ give similar results.

We see that our standard solution excludes the stellar mass range 
($m_2\go0.1\,M_\odot$) for the 
second companion, and this is rather surprising. A stellar mass 
would provide a natural explanation for the anomalously high
eccentricity of the inner binary ($e_1 \approx 0.03$) in terms of secular 
perturbations (Rasio 1994). In addition, a stellar mass would also be 
consistent with a preliminary identification of an optical counterpart 
for the system (Bailyn \etal 1994).
We have seen already that assuming a different value for $\fd_{\rm acc}$
does not change our conclusions. Alternatively, we can try to
vary the value of $\fdddd$ within its fairly large error bar.  
However, we find that varying $\fdddd$ within its formal $1\sigma$ error bar 
still does not produce significant changes in $m_2$. In order to obtain
stellar-mass solutions, we find that we must vary
 $\fdddd$ within a larger $4\sigma$ error bar around the best fit value
$\fdddd_m$ given above. Note that this allows for a change of sign in the
actual value of $\fdddd$. The results are illustrated in Figure~3.
Here we assume that $\sin i_2 = 1$. 
We find that 
stellar-mass solutions (with $m_2 \go 0.1 \rm M_\odot$) are possible only 
if $-0.1 \lo \fdddd/ \fdddd_m \lo 0.1$, i.e., $\fdddd$
must be more than $3\sigma$ away from its present best-fit value
(assuming $\sin i_2 \approx 1$).

\subsection{Secular Perturbations of the Inner Binary}

We can further constrain the system by considering
secular perturbations in the orbital elements of the inner binary.
Preliminary measurements have been made of  the perturbations in
$\omega_1$, $e_1$, and $x_1$ (Arzoumanian \etal 1996; Arzoumanian \& 
Thorsett 1996).
We use these measurements to further constrain the system by
requiring that all our solutions be consistent with these secular 
perturbations.

Since the period $P_2$ of the second companion is much larger than
the period of the inner binary, we can calculate the secular
perturbations assuming that the second companion has a fixed position
in space with respect to the inner binary.
Let $r_{12}$, $\theta_2$ and $\phi_2$ be the fixed spherical polar
coordinates of the second companion, with the origin at the center
of mass of the inner binary, $\phi_2$ measured from pericenter in
the orbital plane, and $\theta_2$ such that $\sin \theta_2 = 1$ for
the coplanar case.
Then the averaged perturbation rates are given by (Rasio 1994)
\begin{eqnarray}
\dot \omega_1 & = & \frac{3 \pi \eta}{P_1} [\sin^2(\theta_2)
(5\cos^2(\phi_2)-1)-1] \\
\dot e_1 & = & \frac{-15 \pi}{2 P_1} \eta e_1 \sin^2(\theta_2) 
\sin(2\phi_2) \\
\dot i_1 & = & \frac{3\pi}{2 P_1} \eta \sin(2\theta_2) 
\cos(\omega_1 + \phi_2)
\end{eqnarray}
where $P_1=16540653\,\rm s$ is the period  of the inner binary, 
$\eta = (m_2/m_1)[(a_p+a_c)/r_{12}]^3$, $a_c$ is the semi-major 
axis of the inner binary companion, and $a_p$ is the semi-major axis
of the pulsar (with respect to the center of mass of the inner binary). 
The projected semi-major axis of the pulsar is $x_p = (1/c)a_p \sin i_1$,
and therefore 
\begin{equation}
\dot x_p = \frac {1}{c} a_p \cos i_1 \dot i_1.
\end{equation}
Note that there is no
secular perturbation of the semi-major axis ($\dot a_p = 0$).

The present measured values of the perturbations are 
(Arzoumanian \& Thorsett 1996)
\begin{eqnarray}
\dot \omega_1 & = & (-2.0\pm2.1)\times10^{-4}\,^{\circ}\,\rm yr^{-1}\\
\dot e_1 & = & (5\pm3)\times10^{-15}\,\rm s^{-1} \\
\dot x_p & = & (-6.5\pm0.8)\times10^{-13}
\end{eqnarray}
Only $\dot x_p$ is clearly detected,
while the two others are at best marginal detections. Note that proper 
motion can also lead to a change in the projected semi-major axis $x_p$ of 
the pulsar. 
However, if the observed $\dot x_p$ were due to  proper motion, 
Arzoumanian \etal (1996) find that the inner
binary companion of the pulsar would then have a mass 
$m_c > 1.0\,\rm M_\odot$, 
with $\sin i_1 < 10^{\circ}$ for a $1.35\,\rm M_\odot$ 
pulsar, which seems very unlikely. Hence we assume here that
the observed $\dot x_p$ is caused  by the secular precession of the 
inner orbit induced by the presence of the second companion.

To incorporate these measurements into our theoretical model,
we have performed Monte-Carlo simulations with the unknown variables in
the system, namely $i_1$, $i_2$, $e_2$, $\theta_2$, and $\phi_2$.
The angles $\theta_2$ and $\phi_2$ can be determined using $i_1$, $i_2$,
$\omega_2$, $\lambda_2$, and an additional undetermined angle $\alpha$,
which (along with $i_1$ and $i_2$) describes the relative 
orientation of the planes of the orbit of the inner binary and the orbit
of the second companion. For a detailed description of the geometry, see
Appendix~A.
We assume a uniform probability distribution for $\cos(i_1)$, $\cos(i_2)$, 
and $\alpha$. Although there is no reason to 
expect the triple system to be in thermal equilibrium with the 
cluster (since the lifetime of the triple system in the core of 
M4 is only $\sim 10^7 \rm\,yr$; see discussion in \S 3.3), 
we assume a thermal distribution 
for $e_2$, i.e., a linear probability distribution Prob($e_2$)$=2 e_2$ (see,
e.g., Heggie 1975) for lack of a better alternative.
Our procedure for constructing random realizations of the system is as
follows.
We start by assuming a value of $e_2$ and solving the nonlinear system 
numerically for $m_2$, $a_2$, $\lambda_2$ and $\omega_2$ as described 
in \S 3.1. Using this solution, we calculate 
$\eta = (m_2/m_1)[(a_p+a_c)/r_{12}]^3$. 
Then, for each trial, we generate random values for $i_1$, $i_2$, 
and $\alpha$, and calculate $\theta_2$ and $\phi_2$ as described in Appendix~A.
We choose the number of trials for each assumed value of $e_2$ so as to 
get a linear distribution for $e_2$ over all the trials. We then 
calculate the secular perturbation rates using equations (33)--(36), 
and we check for consistency with the observed values of the perturbations. 
We use a simple rejection algorithm,  accepting or rejecting a trial 
configuration based on a three-dimensional 
gaussian probability distribution which is the product of  gaussian 
distributions for
$\dot \omega_1$, $\dot e_1$ and $\dot x_p$ centered around  their mean 
values and with standard deviations equal to the error bars given above. 

In Figures~4 and~5, we show histograms of the number of successful
trials for different values of $m_2$ and  the corresponding distance of
the second companion from the inner binary $r_{12}$, for both $\fdddd = 
\fdddd_m$ (standard solutions) and $\fdddd = 0.01 \fdddd_m$ (required 
for obtaining stellar-mass solutions). For 
our standard solutions, we find that the mass distribution for 
the second companion peaks at $m_2 = 0.01\pm0.005\,\rm M_\odot$, 
corresponding to a distance of $r_{12} = 38\pm6\,\rm AU$. Note that
this is the distance of the second companion from the binary
({\emph not} the semi-major axis of the second companion). 
Here the error bars represent 80\% confidence intervals.
The semi-major axis $a_2$ and period $P_2$ are not well constrained 
and vary over three orders of magnitude. The period $P_2$ varies from about
$10^2$ to $10^4$ years with the distribution centered around 400 years,
and $a_2$ varies from about $10$ to $1000\,\rm AU$ with the 
distribution centered around $70\,\rm AU$.
In the second case, the distribution 
peaks at $m_2 = 1.0\pm0.5\,\rm M_\odot$, corresponding to a distance 
of $r_{12} = 150\pm35\,\rm AU$. 
The period $P_2$ varies from about
$10^3$ to $10^5$ years with the distribution centered around 3000 years,
and $a_2$ varies from about $10^2$ to $10^4\,\rm AU$ with the 
distribution centered around $200\,\rm AU$.
We get peaks in the stellar mass 
range  for $m_2$ only when $-0.1 \lo \fdddd/ \fdddd_m \lo 0.1$.
In all cases we find that 
the orientation of the second companion with respect to the inner binary 
(angles $\theta_2$ and $\phi_2$) is poorly constrained by the current
data, as is the inclination $i_2$ of the second companion. The 
inclination of the binary is better constrained to 
$\sim 55^\circ\pm15^\circ$ 
(cf. Fig.~6). The eccentricity of the second companion, $e_2$, is also 
poorly constrained.

\subsection{Model Predictions}

We can obtain a predicted value of the fifth pulse frequency 
derivative $f^{(5)}$ at the current epoch for each of our solutions by 
differentiating equation~(9) for a specified value of $e_2$. 
In Figure~7, we show the most probable values for $f^{(5)}$ from the
Monte-Carlo simulations of \S3.2. For $\fdddd = \fdddd_m$ (standard solution), 
we find that the most probable value is $f^{(5)} =  
(0.15\pm0.05)\times10^{-48}\,\rm s^{-6}$, whereas for 
$\fdddd = 0.01 \fdddd_m$ (which gives stellar mass solutions), 
$f^{(5)} = -6.0\pm0.4\times10^{-51}\,\rm s^{-6}$. 
Thus even a crude measurement of $f^(5)$ 
should completely settle the question of the second companion's mass.

We have also calculated the predicted evolution of the frequency 
derivatives $\fd$ through $\fdddd$ for the next 20 years. We show the
results for the typical orbit of a Jupiter-to-brown-dwarf-size companion
($m_2 = 0.01\,\rm M_\odot$) in Figure~8, and for a stellar-mass companion
($m_2 = 0.5\,\rm M_\odot$) in Figure~9. In the first case, the orbit has
$e_2 = 0.77$, a period $P_2 = 1562\,\rm yr$ and $a_2 = 160\,\rm AU$. We
see that $\fd$ changes sign in about 10 years, and that $\fddd$ decreases
surprisingly fast, changing sign in about 1.5 years.
For the stellar-mass case, only $\fd$ changes sign in 10 years.
The other frequency derivatives do not change significantly over 20 years.
The orbit in this case has $e_2 = 0.49$, a period $P_2 = 2034\,\rm yr$ 
and $a_2 = 161\,\rm AU$. Thus a change in sign of $\fddd$ within the 
next couple of years would provide additional support for the existence 
of a planet or brown dwarf in this system.

We also find that, in all cases, the values of $\fdd$, $\fddd$, and
$\fdddd$ at \emph{apastron} are at least two, three, and five orders of 
magnitude smaller, respectively, than their present observed values. This 
means that the triple nature of the system would probably remain
undetectable near apastron. It is therefore reasonable to find
the second companion relatively close to periastron in our solutions
(within $\sim 15^{\circ}$ for the case illustrated in Fig.~8,
and $\sim 40^{\circ}$ for the case considered in Fig.~9).

\subsection{Discussion}

As mentioned previously, the method for determining the orbital parameters of
a binary pulsar presented in \S 2, although quite general
in its formulation, can only be applied successfully to systems containing
fast millisecond pulsars in which the dynamically-induced 
frequency derivatives dominate the measurements.
parts. In addition, it requires several successive higher-order frequency 
derivatives to be measured accurately. 
The PSR~B1620$-$26 triple system satisfies all of
these conditions, and hence is ideally suited for analysis using our method.
PSR~B1620$-$26 has been observed for more than seven years,
and its hierarchical triple structure is strongly supported by 
all current observations (Backer, Foster, \& Sallmen 1993; 
Thorsett, Arzoumanian, \& Taylor 1993; Arzoumanian \& Thorsett 1996). 
The error bar on $\fdddd$ is likely to shrink 
rapidly as more timing data become available. If the actual value of 
$\fdddd$ is close to the current best-fit value $\fdddd_m = 
-2.1\times10^{-40} \, {\rm s}^{-5}$, then the second companion must 
have a mass $m_2 \leq 0.1\,\rm M_\odot$ as long as the system has an 
inclination $i_2 \geq 7^{\circ}$, with the most probable mass (given 
by our Monte-Carlo simulations) being $0.01\pm0.005\,\rm M_\odot$ (the 
error-bar indicates an 80\% confidence interval). If $\fdddd$ is within 
$1\sigma$ of $\fdddd_m$, then the same result holds to within a factor of 
two. A rather low inclination angle ($i_2 \leq 10^{\circ}$) or
$|\fdddd/\fdddd_m| \leq 0.1$ (i.e., $\fdddd$ more than $3\sigma$ 
away from $\fdddd_m$) would be required if the second companion is a 
main-sequence star with $m_2 \geq 0.1\,\rm M_\odot$.

Instead our results clearly suggest that the second companion is a 
$\sim 0.01\,\rm M_\odot$ brown dwarf or giant planet. This is 
surprising since low-mass objects are not expected to be found in the cores of 
globular clusters. The reason is that low-mass objects have higher velocities in energy
equipartition and are preferentially ejected 
from globular clusters as they evaporate in the tidal field of the Galaxy.
Hence we do not expect to find very low-mass stars or brown dwarfs
in globular clusters, especially not near the core. Recent 
HST observations of globular clusters (e.g., Paresce, De Marchi,
\& Romaniello 1995) also support 
this view by finding that stellar mass functions in clusters flatten or even
drop for masses below $\sim 0.1\,\rm M_\odot$. 
In addition, if the second companion of PSR~B1620$-$26
is indeed of low mass, then the
unusually high eccentricity of the inner binary pulsar cannot be explained 
by secular perturbations due to the second companion, since that would 
require a stellar-mass second companion (Rasio 1994). 
It would also 
preclude any possibility of an optical identification of the triple
system. Bailyn \etal (1994) have searched deep optical images of 
M4 for an optical counterpart of the pulsar. 
They have identified  a candidate which, if interpreted as a single 
object, could be a $0.45\,\rm M_\odot$ main-sequence star within 
$0.3\arcsec$ of the nominal pulsar position. However, it is possible
that this object is in fact a blend of fainter stars not associated 
with the pulsar, or simply a chance superposition. Future observations 
of the region with HST, as well as improved 
ground-based astrometry, should help resolve the issue. 

Low-mass stars and brown dwarfs could exist in dense globular cluster cores as
binary companions to more massive stars. Dynamical interactions could then lead
to an exchange, leaving the low-mass object in orbit around a neutron star. 
Indeed, Sigurdsson (1992) had discussed the possibility of finding planetary 
companions to pulsars in globular clusters even before the triple nature of 
PSR~B1620$-$26 was established.
A possible formation scenario for the triple system starts with an interaction 
between a neutron-star-white-dwarf binary  and a main-sequence star with a large 
Jupiter-type planet or brown-dwarf companion (cf.\ Sigurdsson 1995, 1993). 
As a result of this interaction the white dwarf is ejected while the 
main-sequence star and its planet or brown-dwarf companion 
remain in orbit around the neutron star. 
The main-sequence star, as it evolves and later expands as
a red giant, would then 
transfer mass onto the neutron star, thus spinning it up and forming 
the millisecond pulsar in the triple configuration we see today. 
However, tidal dissipation during the mass transfer 
phase would effectively circularize 
the orbit of the binary, leaving a residual eccentricity 
$e_1 \lo 10^{-4}$ (Phinney 1992). Therefore this formation
scenario leaves the much higher observed eccentricity ($e_1 \approx 0.03$)
of the inner binary unexplained.
It has been suggested that the eccentricity of the inner binary may have been 
induced during a dynamical interaction with another cluster star. 
The probability of disrupting the triple during such an interaction is only 
$\sim 0.5$ (Sigurdsson 1995). Based on the results of Rasio \& Heggie (1995), 
however, 
we find that the observed eccentricity would require an encounter with a
distance of closest approach of $\sim 2.5\,\rm AU$, considerably smaller 
than the size of the outer orbit, and occuring on average once in
$\sim 4\times10^{8}\,\rm yr$.
For comparison,
the lifetime of the triple system in the cluster is only about
$\tau \sim 10^8{\,\rm yr}\,\rho_4^{-1} \sigma_5 (a_2/10{\,\rm AU})^{-1}
\sim 2\times10^7{\,\rm yr}$,
where   
$\rho = 10^4\rho_4\,\rm M_\odot\,pc^{-3}$ is the density near the 
center of M4, $\sigma = 5 \sigma_5\,\rm km\,s^{-1}$ is the velocity 
dispersion, and $a_2$ is the size of the outer orbit (Rasio 1994). 
For one interaction that could have produced the eccentricity of the 
inner binary,
we therefore expect $\sim 20$ interactions that could have disrupted the 
triple,
each with probability $\sim 0.5$, leaving the probability of survival at 
$\sim 10^{-6}$.
An additional problem is that the age of the millisecond pulsar in this scenario
must be comparable to the age of the triple ($\sim10^7\,\rm yr$), which 
requires the millisecond pulsar to be extremely young.
This problem could be avoided if the triple was instead formed during an interaction 
involving a {\em pre-existing\/} binary millisecond pulsar and another primordial
binary (containing the present second companion and another star that was ejected during
the interaction; see Rasio, McMillan, \& Hut 1995). 
The current eccentricity of the (inner) binary pulsar could then have been
induced during the same interaction that formed the triple, although this would 
require some fine tuning. More significantly,
one would expect the more massive member of the other binary 
rather than the low-mass object (Jupiter or brown dwarf) 
to be preferentially retained in the triple while the other gets ejected.

Naturally, if the low-mass object (Jupiter or brown dwarf) was attached
to a much more massive star, it is easier to understand how it was retained 
by the cluster and why it is now found close to the cluster core.
In particular, in the first formation scenario discussed above, the 
main-sequence star must have been fairly massive ($m\sim1\,\rm M_\odot$) 
to have evolved into a red giant after the triple was formed. Two-body 
relaxation in the cluster will tend to bring this main-sequence
star (with its attached low-mass companion) down to the cluster core since 
it is more massive than the average object in the cluster.
Confirmation of the existence of a $\sim 0.01\,\rm M_\odot$ object in 
PSR~B1620$-$26 would therefore provide further 
indication that many stars, even in globular clusters, could 
have very low-mass companions or planets. This is especially important in 
the light of recent discoveries of several 
$\sim 10^{-3}-10^{-2}\,\rm M_\odot$ objects around nearby 
stars (e.g., Mayor \& Queloz 1995; Butler \& Marcy 1996; 
Marcy \& Butler 1996).

\section{Application to the PSR~B1257$+$12 Planetary System}

We now turn to the application of our method to the planetary system around 
the millisecond pulsar PSR~B1257$+$12. This system contains three 
confirmed Earth-mass planets in quasi-circular orbits 
(Wolszczan \& Frail 1992; Wolszczan 1994).  
The planets have masses of $0.015/\sin i_1\,\rm M_\oplus$, 
$3.4/\sin i_2\,\rm M_\oplus$, and $2.8/\sin i_3\,\rm M_\oplus$, where $i_1$, 
$i_2$ and $i_3$ are the inclinations of the orbits with respect to the 
line of sight, 
and are at distances of 0.19\,AU, 0.36\,AU, and 0.47\,AU, respectively, 
from the pulsar. In addition, the unusually large second and third frequency 
derivatives of the pulsar suggest the existence of a fourth, more distant 
and massive planet in the system (Wolszczan 1996).

\subsection{Analysis of the Frequency Derivative Data}

The residual pulse frequency derivatives for PSR~B1257$+$12 (after 
subtraction of a model for the inner three planets) are 
$\fd  = -8.6\times10^{-16}\,\rm s^{-2}$, 
$\fdd  = (-1.25\pm0.05)\times10^{-25}\,\rm s^{-3}$, and 
$\fddd = (1.1\pm0.3)\times10^{-33}\,\rm s^{-4}$ (Wolszczan 1996),
while the frequency $f = 160.8\,\rm s^{-1}$. The value 
of $\fd$ has been corrected for the apparent 
acceleration due to the pulsar's transverse velocity 
(the so-called Shlovskii effect; cf.\ Camilo \etal 1994). 
The error bars on $f$ and $\fd$, for the purposes of this 
discussion, are negligible. Note that the measurement of $\fddd$ is
only preliminary, but we assume here that the value quoted above 
(from Wolszczan 1996) is correct.
Comparison with PSR~B1855$+$09, which has a very similar pulse frequency, 
$f = 186\,\rm s^{-1}$, and first frequency derivative
$\fd = -6.2\times10^{-16}\,\rm s^{-2}$ (Kaspi \etal 1994) 
indicates that the observed $\fd$ for PSR~B1257$+$12 could well be 
entirely (or in large part) intrinsic rather than acceleration-induced.
The timing age for the pulsar, $\tau = -f/2\fd \sim 3\times10^9\,\rm yr$,
is entirely consistent with that expected for a millisecond pulsar.
Therefore we will treat $\fd_{\rm acc}$ essentially as a free parameter in our
analysis.
The observed $\fdd$, on the other hand, is two orders of magnitude
larger than for PSR~B1855$+$09, which has 
$\fdd \leq 2.0\times10^{-27}\,\rm s^{-2}$. 
Thus the observed $\fdd$ is almost certainly 
due to the presence of another planet rather than intrinsic
timing noise in the pulsar.

With three frequency derivatives measured, we can use the method of \S 2.3
to model the system. Given the near-circular 
orbits of the three inner planets, it is natural to assume that the orbit 
of the fourth planet also has a low eccentricity. In addition, it is easy 
to show that dynamical interactions with passing stars in the Galaxy 
are not likely to produce any
significant perturbations of the system (which could otherwise increase 
the eccentricity of an outer planet's orbit; cf. Heggie \& Rasio 1996).

Since the value of $\fd_{\rm acc}$ is uncertain, 
we explore a wide range, $0.01<\fd_{\rm acc}/\fd_{\rm obs}<1$.
Note that, for a circular orbit, $\fd_{\rm acc}$ and $\fddd$ must have opposite
signs (cf.\ eqs. [19] and [21]). Hence $\fd_{\rm acc}$ cannot be positive.
For each value of 
$\fd_{\rm acc}$, we calculate the mass and semi-major axis of the fourth 
planet using equations~(25), (28) and~(30). We illustrate the results in 
Figure~10. 
We find that the mass of the fourth planet varies significantly, from 
$\sim 0.08\,\rm M_\oplus$ (for $\fd_{\rm acc} = 0.01 \fd_{\rm obs}$) 
to $\sim 100\,\rm M_\oplus$ (for $\fd_{\rm acc} = \fd_{\rm obs}$). 
The simplest interpretation of the 
present best-fit values of the frequency derivatives, assuming
$\fd_{\rm acc} = \fd_{\rm obs}$, implies 
a mass of about $100/\sin i_4 \,\rm M_\oplus$ (i.e., comparable to 
Saturn's mass) for the fourth planet, at a distance of about
$38\,\rm AU$ (i.e., comparable to Pluto's distance from the Sun), and
with a period of about $170\,\rm yr$ in a circular, coplanar orbit 
(Wolszczan 1996). 
However, if $\fd_{\rm acc} \neq \fd_{\rm obs}$, 
then the fourth planet can have a wide range of masses.
In particular, it can have a mass comparable to that of 
Mars (at a distance of 9 AU), Uranus (at a distance of 25 AU) or 
Neptune (at a distance of 26 AU),  
for $\fd_{\rm acc} = 0.015$, 0.30, or $0.34\,\fd_{\rm obs},$ 
respectively.

\subsection{Discussion}

In this system the perturbations of the inner planets produced by the 
fourth planet are probably far too small to be detected. 
This is in contrast to the mutual perturbations of the inner planets
themselves, which are important and have been detected
(Rasio et al.\ 1992; Wolszczan 1994). Using equations
(33)--(36), we predict $\dot e \sim 10^{-17}\,\rm s^{-1}$ and 
$\dot \omega \sim 10^{-7}\,\rm deg\,yr^{-1}$ for the orbit of the
third planet, assuming that all orbits are coplanar and that the mass of
the fourth planet is $100\,\rm M_\oplus$. 
The perturbations for the two innermost planets are even smaller. 
Hence the existence of 
the fourth planet is likely to be confirmed only through further
measurements of pulse frequency derivatives.

It has been pointed out that the masses and radii of the 
three inner planets in PSR~B1257$+$12 are in the same ratios as the 
masses and radii of the corresponding first three planets in the 
Solar System (Mazeh \& Goldman 1995). This might perhaps 
be indicative of a global underlying formation mechanism for the 
two systems. 

Although the fourth planet could have the same mass (normalized to the 
mass of the third planet) as that of 
Mars, Uranus, or Neptune (normalized to the mass of the Earth), 
the ratio of radii in each case would be much 
larger than the corresponding ratio for the solar system (cf. Fig.~10). 
Thus this system does not seem
to maintain its regularity with the Solar System, since the mass and radius 
ratios of the fourth planet would not simultaneously match those of 
any planet in the Solar System.
This is true for the entire range of values of $\fd_{\rm acc}$ 
considered above. 

\newpage

\acknowledgements

We are very grateful to Z.~Arzoumanian, S.~E.\ Thorsett and A.\ Wolszczan 
for many useful discussions and for communicating results of observations 
in progress. We also thank C.\ Bailyn and S.\ Sigurdsson for helpful
comments.
F.A.R.\ is supported by an Alfred P.\ Sloan Research Fellowship.

\appendix
\section{Geometry of the Triple Configuration}

The orbit of the inner binary and the orbit of the second companion in general
do not lie in the same plane. The inclinations of the two planes with respect to
the line of sight are given by $i_1$ and $i_2$. To specify the plane of an orbit
completely, one needs the inclination angle together with another azimuthal angle 
$\alpha$ (which lies between 0 and $2\pi$). Since the reference axis for $\alpha$
is arbitrary, we can take it to lie in the plane of one of the orbits, so that 
$\alpha$ is the difference between the azimuthal angles of the two planes. 
In random Monte-Carlo trials, $\alpha$ is then taken to be uniformly distributed 
between $0$ and $2\pi$. 

In order to determine $\theta_2$ and $\phi_2$ using the other angles, we need
to change coordinates between two reference frames. The first frame has its origin
at the center of mass of the inner binary, with the x-axis in the plane of the 
orbit of the second companion, the y-axis passing through the pericenter of the 
orbit, 
and the z-axis perpendicular to the plane of the orbit, so that the motion of 
second companion is counterclockwise around the z-axis. We shall refer to the 
coordinates of the second companion in this frame as ($x_c^{\prime}$, 
$y_c^{\prime}$, $z_c^{\prime}$). Then, $x_c^{\prime} = -r_{12} \sin \lambda_2$,
$y_c^{\prime} = r_{12} \cos \lambda_2$, and $z_c^{\prime} = 0$.

The second frame similarly has its origin at the center of mass of the inner binary, 
with the x-axis in the plane of the orbit of 
the inner binary, the y-axis passing through the pericenter of the orbit, 
and the z-axis perpendicular to the plane of the orbit, so that the motion of 
pulsar is counterclockwise around the z-axis. We wish to find the 
coordinates ($x_c^{\prime\prime}$, $y_c^{\prime\prime}$, $z_c^{\prime\prime}$) 
of the second companion in this frame. We can then calculate $\theta_2$ and 
$\phi_2$ using the formulae $\theta_2 = \cos^{-1}(z_c^{\prime\prime}/r_{12})$ 
and $\phi_2 = \tan^{-1}(-x_c^{\prime\prime}/y_c^{\prime\prime})$, keeping in 
mind that for $y_c^{\prime\prime} < 0$, we must add $180^{\circ}$ to 
$\phi_2$ in order to get the correct quadrant.

In order to get ($x_c^{\prime\prime}$, $y_c^{\prime\prime}$, $z_c^{\prime\prime}$)
from ($x_c^{\prime}$, $y_c^{\prime}$, $z_c^{\prime}$), we shall rotate the first 
frame to the second frame using the standard Euler angles formalism (see, e.g.,
Goldstein 1980). In this formalism, any arbitrary rotation of an object is 
represented as a sequence of three consecutive rotations- first about the z-axis
by an angle $\phi$, then about the new x-axis by an angle $\theta$, and finally
again about the new z-axis by an angle $\psi$. 

In order to use this formalism, we will use an intermediate frame of reference 
that is fixed in space, with its origin at the center of mass of the inner binary, 
the y-axis along the line of sight, the x-axis in the plane of the orbit of the
second companion, and the z-axis such that the motion of the second companion 
is counterclockwise about the z-axis. The first frame described above can be 
obtained from this fixed frame by rotating it through the Euler angles
$0$, $i_2$ and $\omega_2 - 90^{\circ}$, respectively. Similarly, the second frame
can be obtained from the fixed frame by rotating it through the angle $\alpha$
about the y-axis, and then rotating it through the Euler angles 
$0$, $i_1$ and $\omega_1 - 90^{\circ}$, respectively.
The sequence of Euler angle rotations is represented as a matrix 
${\bf\rm A}(\phi, \theta, \psi)$. 
We get the coordinates 
($x_c^{\prime\prime}$, $y_c^{\prime\prime}$, $z_c^{\prime\prime}$) 
from ($x_c^{\prime}$, $y_c^{\prime}$, $z_c^{\prime}$) by multiplying
first by the inverse matrix ${\bf\rm A}^{-1}(0, i_2, \omega_2 - 90^{\circ})$,
then by multiplying by the matrix for rotation about the y-axis 
${\bf\rm B}(\alpha)$, and then
multiplying by the matrix ${\bf\rm A}(0, i_1, \omega_1 - 90^{\circ})$.
The matrices are given by,

\[ {\bf{\rm A}}(\phi, \theta, \psi) = 
 \left( \begin{array}{clcr} 
\cos\psi \cos\phi - \cos\theta \sin\phi \sin\psi & 
\cos\psi \sin\phi + \cos\theta \cos\phi \sin\psi & \sin\psi \sin\theta \\
-\sin\psi \cos\phi - \cos\theta \sin\phi \cos\psi & -\sin\psi \sin\phi + 
\cos\theta \cos\phi \cos\psi & \cos\psi \sin\theta \\
\sin\theta \sin\phi & -\sin\theta \cos\phi & \cos\theta       
\end{array} \right) \]

\[ \rm{and} \hspace{0.5in}{\bf{\rm B}}(\alpha) = 
 \left( \begin{array}{clcr} 
	\cos\alpha &  0  &  -\sin\alpha \\
	0  &  1  &  0 \\
	\sin\alpha & 0 & \cos\alpha 
\end{array}  \right). \]

\clearpage

\begin{figure}
\plotone{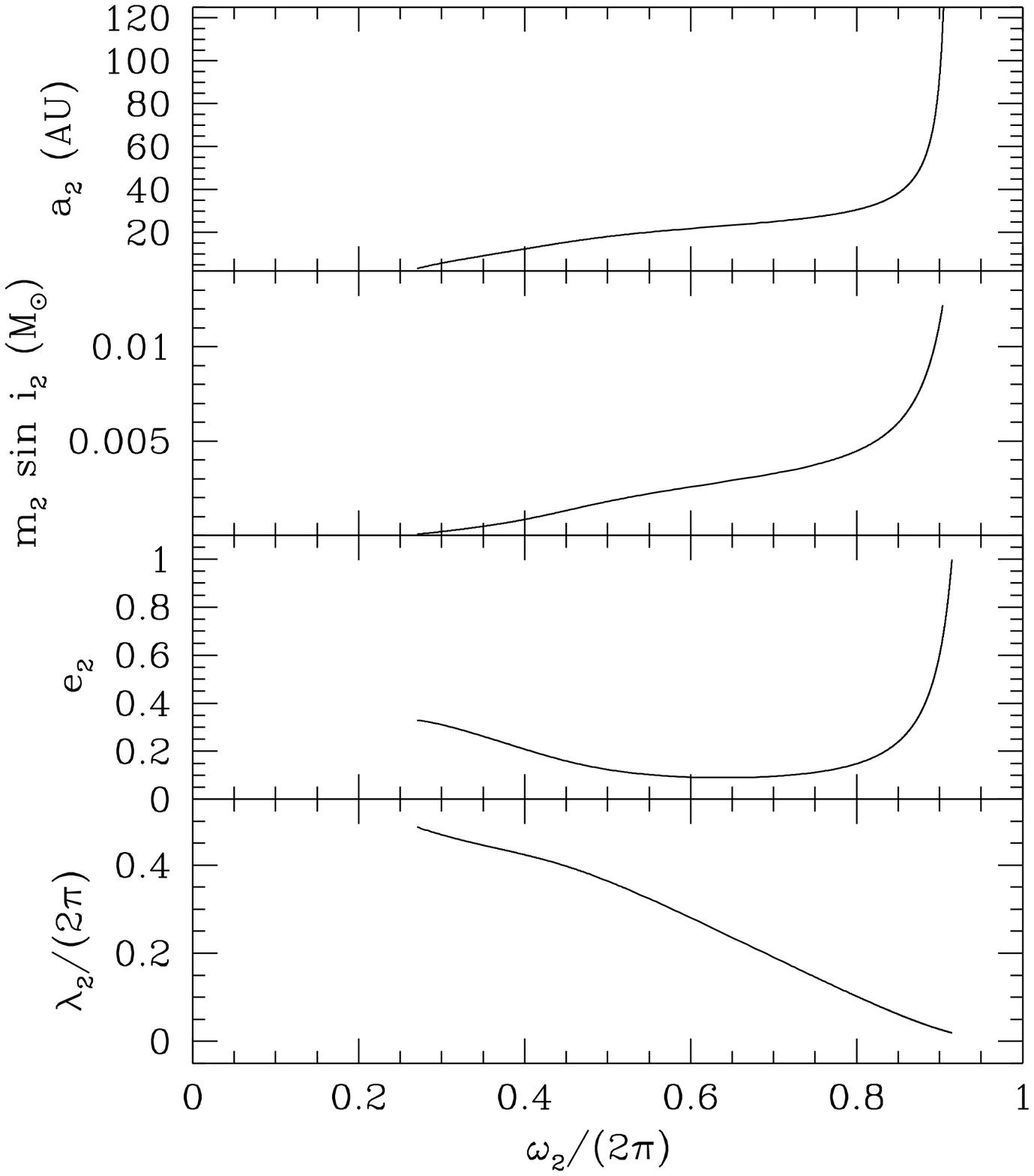}
\caption{Allowed values of the semi-major axis $a_2$, mass $m_2$,
eccentricity $e_2$, longitude at epoch $\lambda_2$, and longitude
of pericenter $\omega_2$
for the second companion of PSR~B1620$-$26, using the latest 
available values for all pulse frequency derivatives. This is our 
``standard solution,'' using the present best-fit value 
$\fdddd_m = -2.1\times10^{-40} \,\rm s^{-5}$ and assuming that all 
measured frequency derivatives are dynamically induced.
Acceptable solutions all have $2.4\times10^{-4}\,\rm M_\odot \leq m_2 
\sin i_2 \leq 1.2\times10^{-2}\,\rm M_\odot$.
\label{fig1}}
\end{figure}

\begin{figure}
\plotone{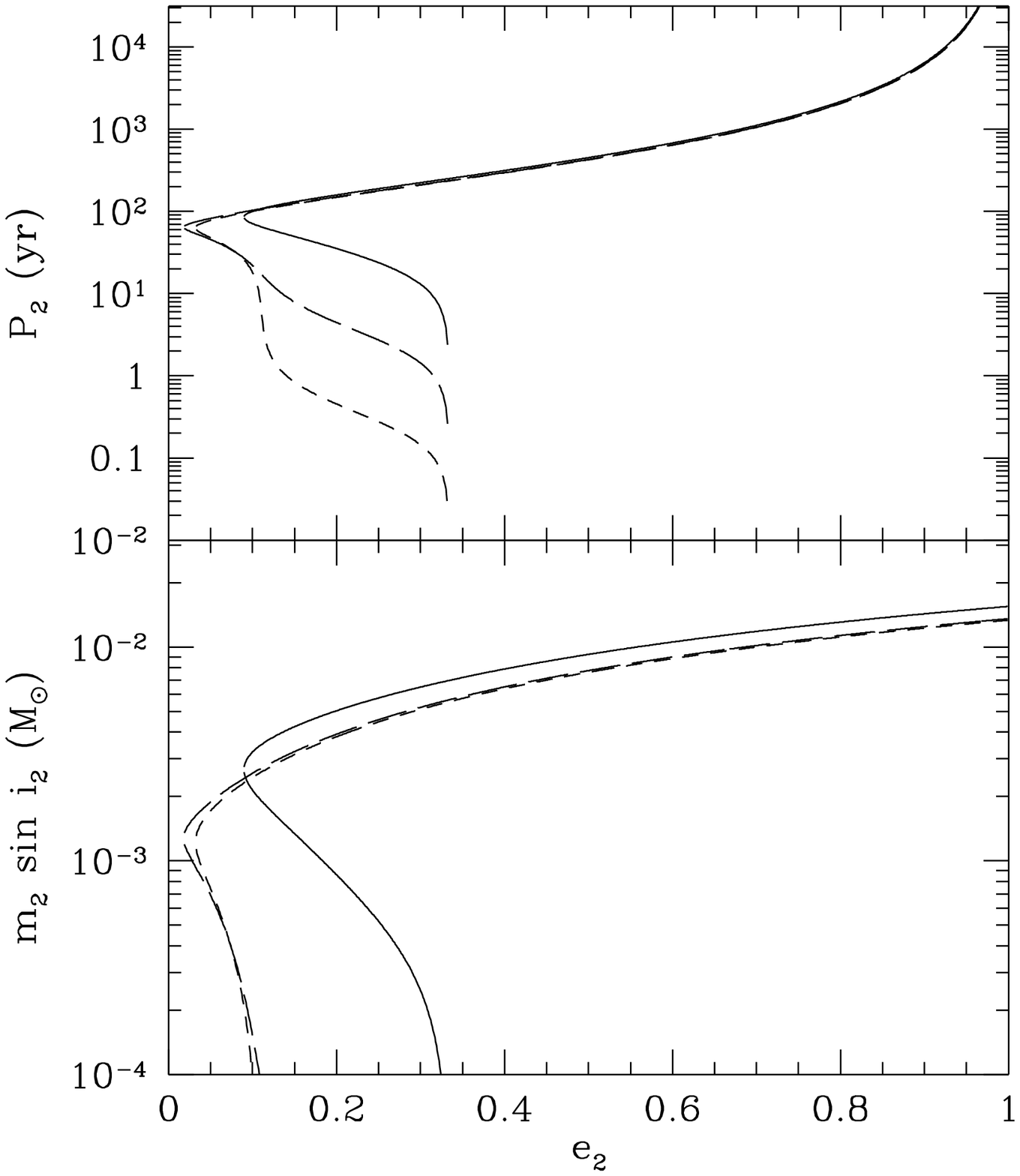}
\caption{The orbital period, mass and eccentricity of the second 
companion of
PSR~B1620$-$26, for different values of the acceleration-induced 
first frequency derivative. 
The various curves are for $\dot f_{\rm acc} = \dot f_{\rm obs}$ 
(solid line), $\dot f_{\rm acc} = 0.1\dot f_{obs}$ (long-dashed line), 
and $\dot f_{\rm acc} = 0.01\dot f_{obs}$ (short-dashed line).
We assume that the inclination angle $i_2=90^\circ$. We see that the
mass range does not change significantly when varying $\dot f_{\rm acc}$, 
and stellar masses are always excluded.
\label{fig2}}
\end{figure}

\begin{figure}
\plotone{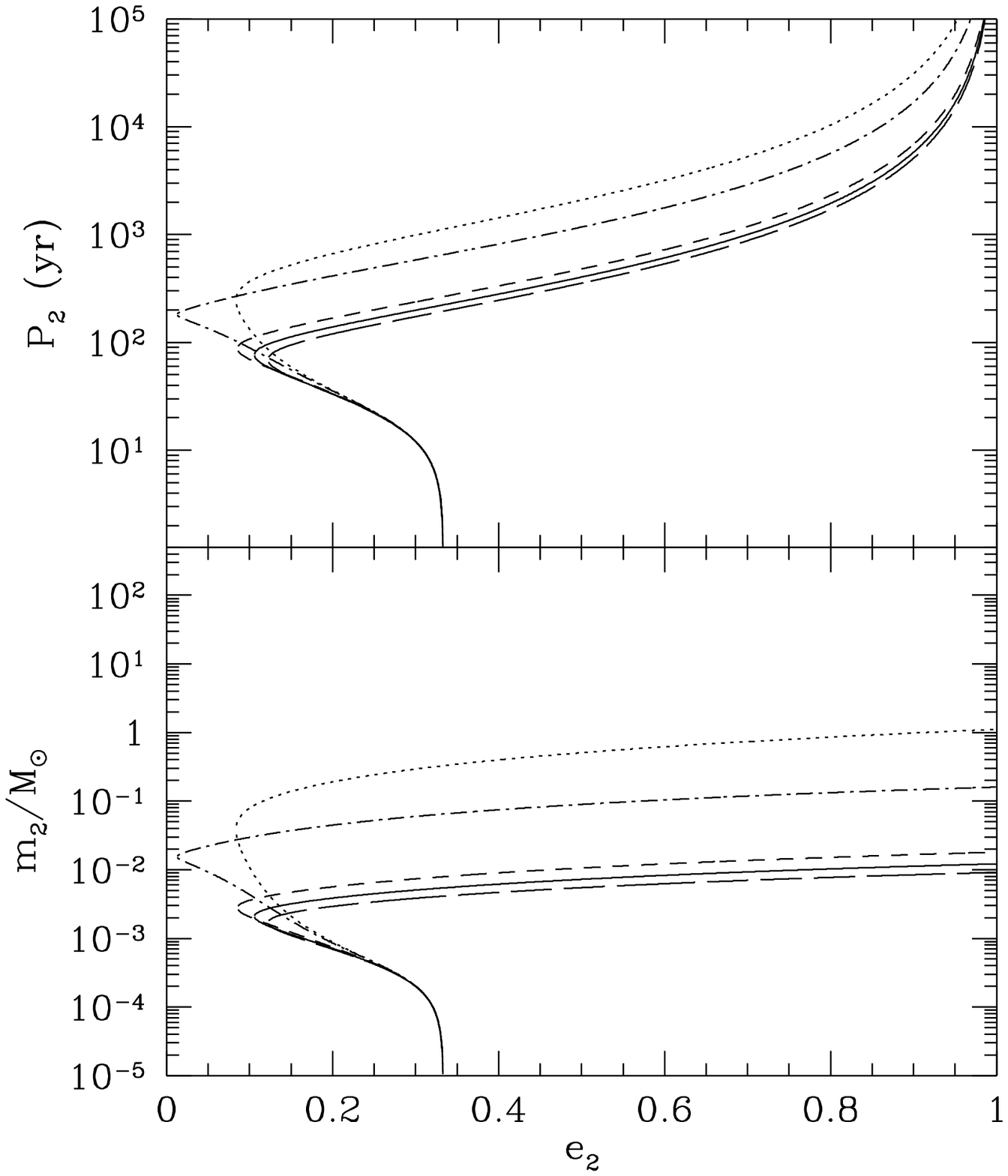}
\caption{Similar to Fig.~2, but the dependence of the solutions on 
$\fdddd$ is illustrated for an expanded $4\sigma$ error bar around 
the best-fit
value $\fdddd_m = -2.1\times10^{-40} \,\rm s^{-5}$. The 
solutions are shown for $\fdddd = \fdddd_m$ (solid line), 
$\fdddd = \fdddd_m - 1\sigma$ (long-dashed line), 
$\fdddd = \fdddd_m + 1\sigma$ (short-dashed line),
$\fdddd = 0.1 \fdddd_m$ (dot-dashed line), and 
$\fdddd = 0.01 \fdddd_m$ (dotted line), 
where $\sigma = 0.6\times10^{-40} \,\rm s^{-5}$. Stellar-mass solutions
are obtained when $-0.1 \lo \fdddd/ \fdddd_m \lo 0.1$
\label{fig3}}
\end{figure}

\begin{figure}
\plotone{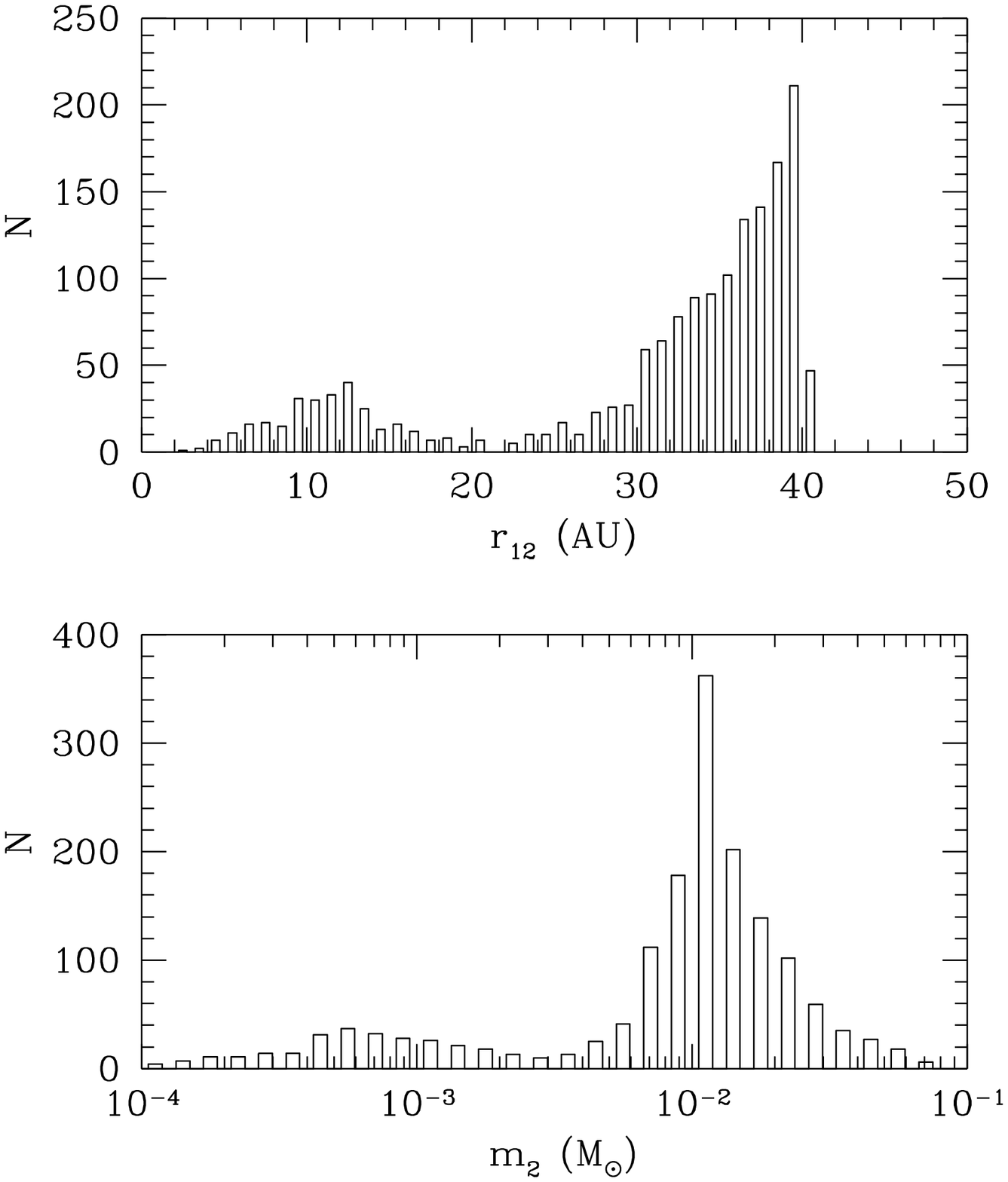}
\caption{Histogram of the number of successful  trials (N) for
different values of $m_2$ and  the corresponding distance of  the
second companion from the inner binary $r_{12}$, for  the case
$\fdddd = \fdddd_m$ in our Monte-Carlo simulations. 
We find that the most probable value for the
second companion mass is $m_2 = 0.01\pm0.005\,\rm M_\odot$, 
corresponding to a distance of $r_{12} = 38\pm6\,\rm AU$ (80\% confidence
intervals).
\label{fig4}}
\end{figure}

\begin{figure}
\plotone{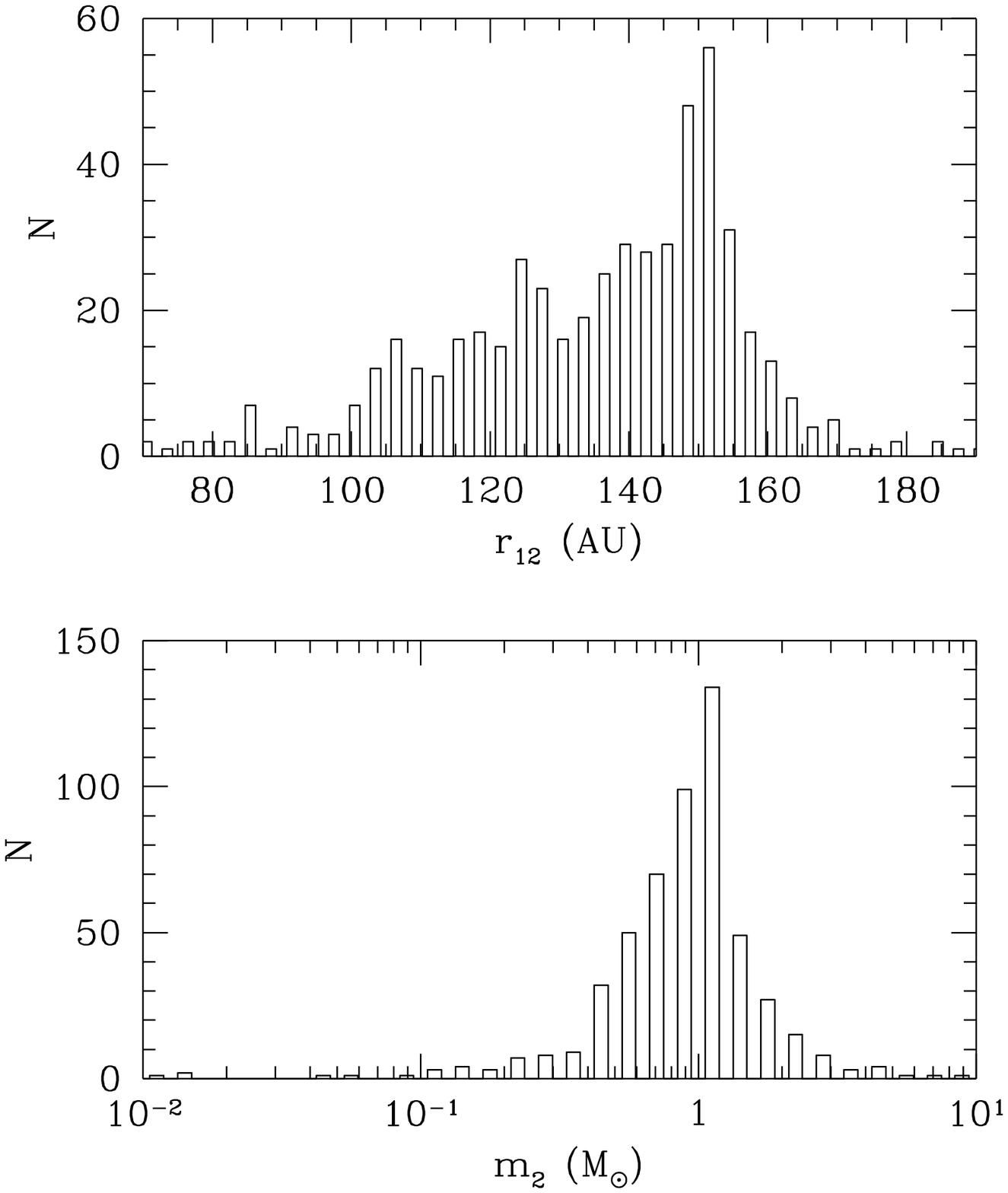}
\caption{Same as Fig.~4, but for the case 
$\fdddd = 0.01\fdddd_m$. We see that the most probable value 
for the second companion mass is now $m_2 = 1.0\pm0.5\,\rm M_\odot$, 
corresponding to a distance of $r_{12} = 150\pm35\,\rm AU$.
\label{fig5}}
\end{figure}

\begin{figure}
\plotone{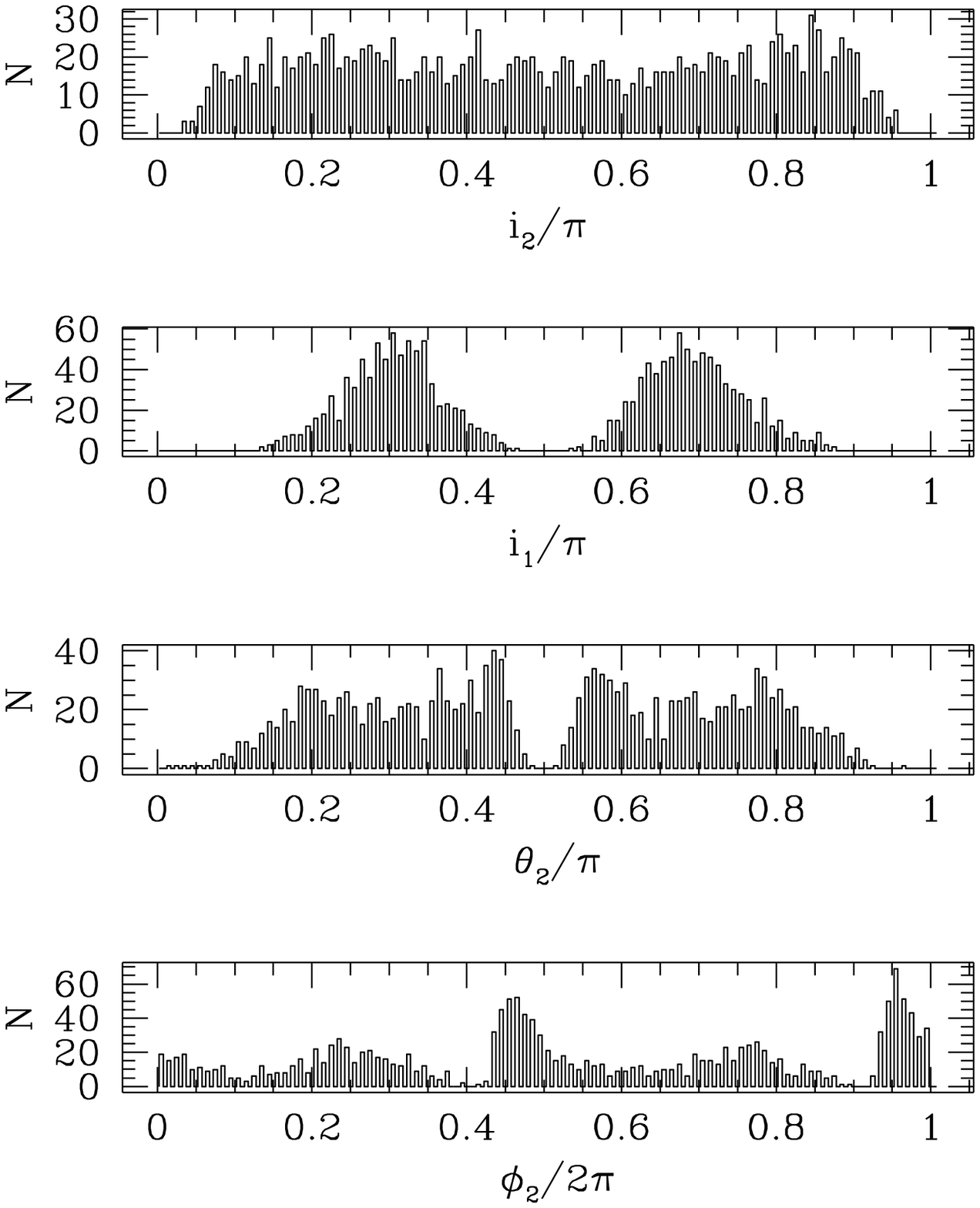}
\caption{Histograms of the number of successful  trials (N) for 
various parameters in the
Monte-Carlo simulations with $\fdddd = \fdddd_m$. All the angles are 
poorly constrained except the inclination of the inner binary, which 
is slightly better constrained to $i_1 \sim (55\pm15)^\circ$.
\label{fig6}}
\end{figure}

\begin{figure}
\plotone{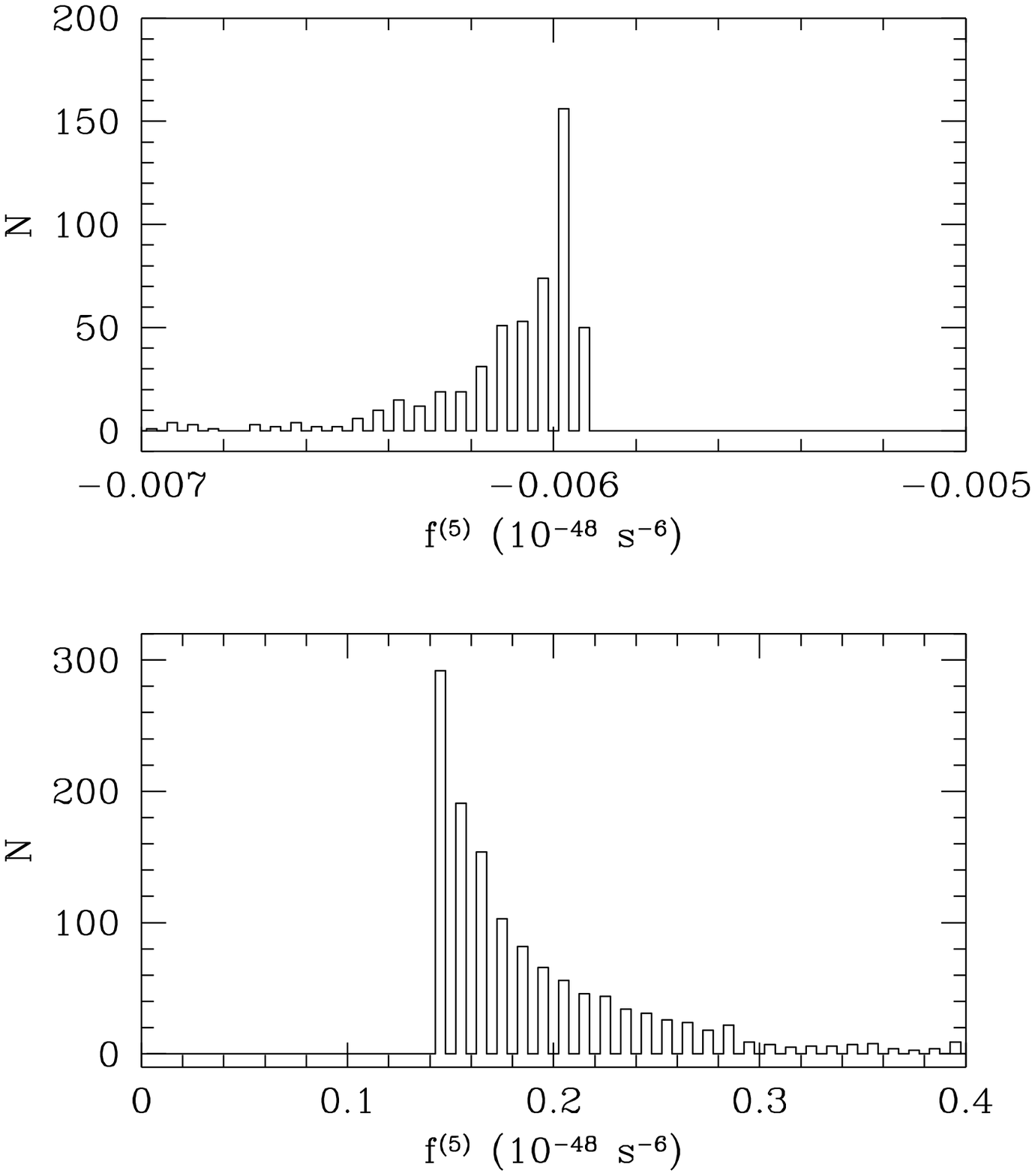}
\caption{The most probable value of the fifth frequency derivative 
$f^{(5)}$ given by the Monte-Carlo 
simulations. For $\fdddd = \fdddd_m$ (lower frame), 
$f^{(5)} \approx 0.15(5)\times10^{-48}\,\rm s^{-6}$ and for 
$\fdddd = 0.01 \fdddd_m$ (upper frame), $f^{(5)} \approx 
-0.0060(4)\times10^{-48}\,\rm s^{-6}$.
\label{fig7}}
\end{figure}

\begin{figure}
\plotone{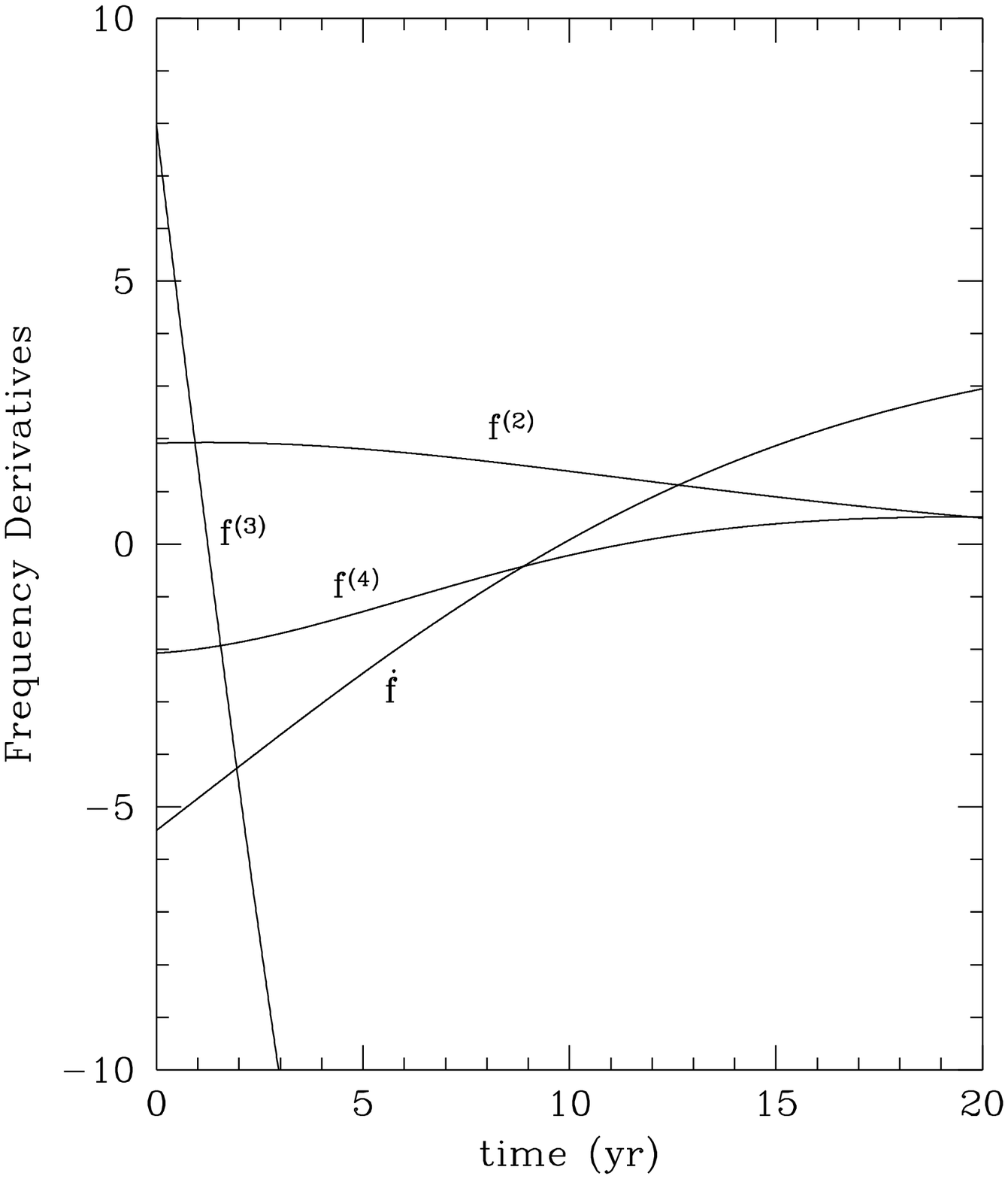}
\caption{
The predicted variation of the frequency derivatives over the
next 20 years, for a low-mass ($m_2 = 0.01\,\rm M_\odot$)
second companion (standard solution). The units for the four
derivatives are $10^{-15}\,\rm s^{-2}$ for $\fd$, 
$10^{-23}\,\rm s^{-3}$ for $\fdd$, $10^{-33}\,\rm s^{-4}$ for $\fddd$, 
and $10^{-40}\,\rm s^{-5}$ for $\fdddd$. We see that $\fd$ changes 
sign in about 10 years, and $\fddd$ changes sign in about 1.5 years.
\label{fig8}}
\end{figure}

\begin{figure}
\plotone{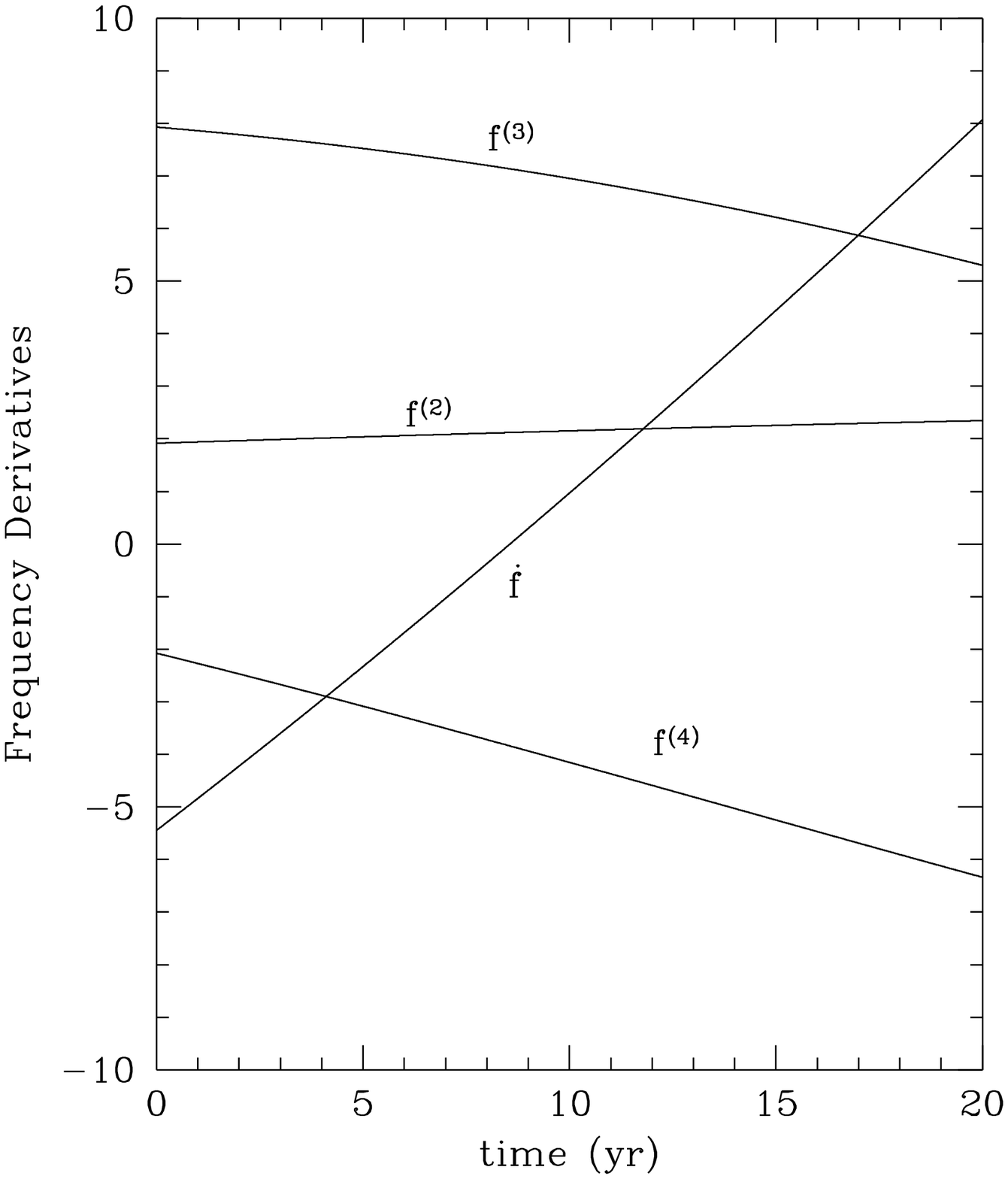}
\caption{The predicted variation of the frequency derivatives over the
next 20 years, for a stellar-mass ($m_2 = 0.5\,\rm M_\odot$)
second companion (assuming $\fdddd = 0.01 \fdddd_m$). The units for the four
derivatives are $10^{-15}\,\rm s^{-2}$ for $\fd$, 
$10^{-23}\,\rm s^{-3}$ for $\fdd$, $10^{-33}\,\rm s^{-4}$ for $\fddd$, 
and $10^{-42}\,\rm s^{-5}$ for $\fdddd$. We see that $\fd$ changes 
sign in about 10 years, but the other derivatives do not change much 
in 20 years.
\label{fig9}}
\end{figure}

\begin{figure}
\plotone{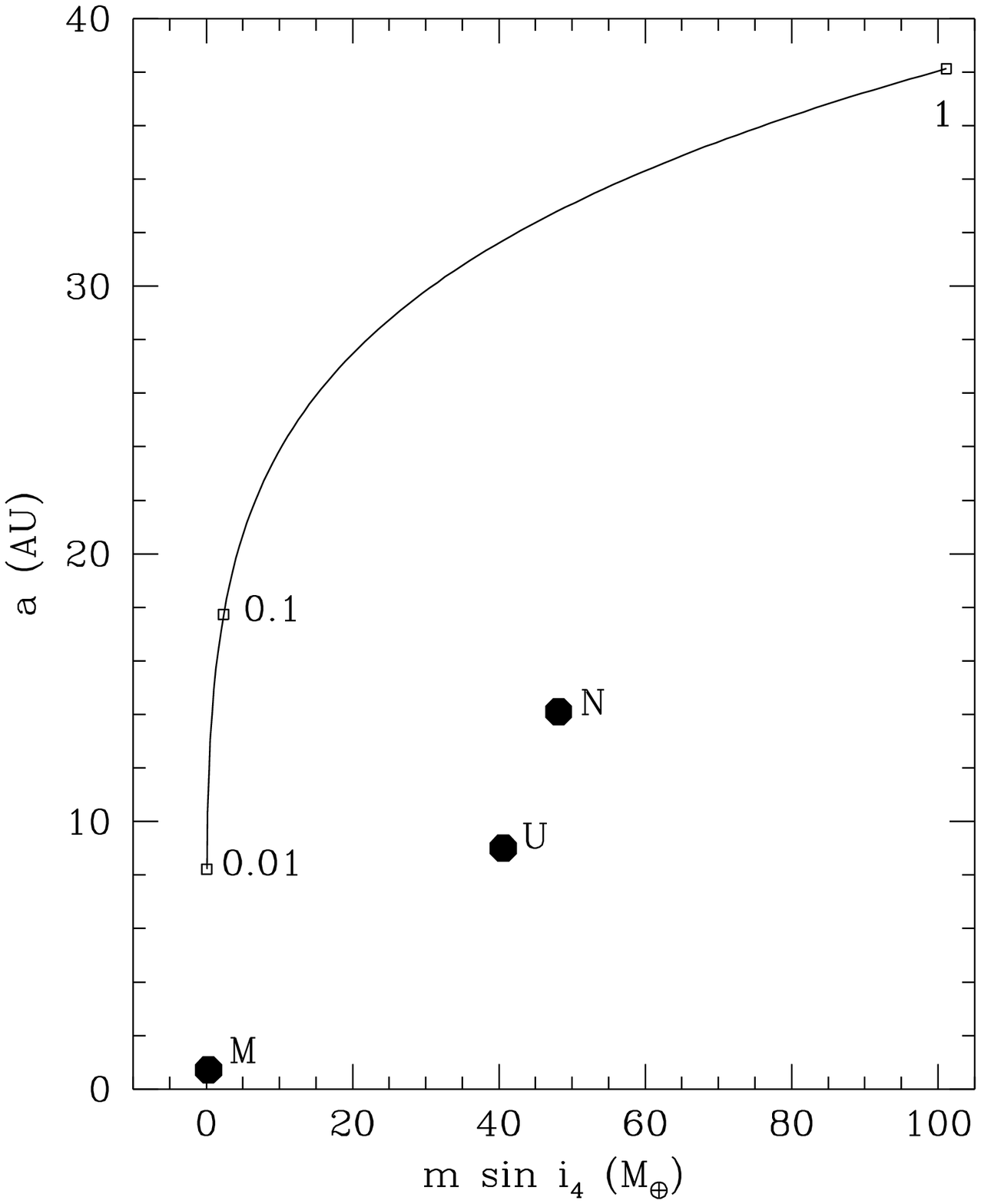}
\caption{The mass and semi-major axis of the possible outer planet 
in the PSR~B1257$+$12 planetary system for $\fd_{\rm acc}$ in the 
range 0.01--1.0 $\fd_{\rm obs}$.
The present best-fit values of the frequency derivatives 
with $\fd_{\rm acc} = \fd_{\rm obs}$ imply the presence of a planet  
with mass $\approx 100/\sin i_4 \,\rm M_\oplus$, at a distance of  
$\approx 38\,\rm AU$. The marked points on the curve indicate the 
values of $\fd_{\rm acc}/\fd_{\rm obs}$. The points labelled M, U and
N indicate configurations with the same mass and radius ratios (in this 
system) as those of Mars, Uranus and Neptune (in the Solar System), 
respectively. 
\label{fig10}}
\end{figure}

\end{document}